\documentclass[twocolumn,amsmath,amssymb,superscriptaddress]{revtex4}
 \usepackage{amsmath} 

\usepackage{float}
\usepackage{graphicx}
\usepackage{dcolumn}
\usepackage{bm}

\newcommand{\nl}{\nonumber \\ &&}
\newcommand{\er}[1]{(\ref{#1})}

\begin{document}
%
\def\ih{\frac{i}{\hbar}}
\def\be{\begin{equation}}
\def\ee{\end{equation}}
\def\bea{\begin{eqnarray}}
\def\eea{\end{eqnarray}}
\def\ds{\displaystyle}
\def\bra#1{\mbox{$\langle#1|$}}
\def\ket#1{\mbox{$|#1\rangle$}}
\def\braket#1{\mbox{$<#1>$}}
\def\pd{\displaystyle\frac{\partial}{\partial t}}
\def\rso{{\hat \rho}}
%
%
%
\title{Calculating optical absorption spectra of thin polycrystalline films:\\ Structural disorder and site-dependent van der Waals interaction}
\author{J\"org Megow}
\email{megow@uni-potsdam.de}
\affiliation{Institut f\"ur Chemie,
Universit\"at Potsdam,
Karl-Liebknecht-Stra\ss{}e 24-25,
D-14476 Potsdam, F. R. Germany}
\author{Thomas K\"orzd\"orfer}
\affiliation{Institut f\"ur Chemie,
Universit\"at Potsdam,
Karl-Liebknecht-Stra\ss{}e 24-25,
D-14476 Potsdam, F. R. Germany}

\author{Thomas Renger}
\affiliation{Institut f\"ur Theoretische Physik,
Johannes Kepler Universit\"at Linz,
Altenberger Stra\ss{}e 69,
A-4040 Linz, Austria}

\author{Mino Sparenberg}
\affiliation{Institut f\"ur Physik,
Humboldt--Universit\"at zu Berlin,
Newtonstra\ss{}e 15,
D-12489 Berlin, F. R. Germany}

\author{Sylke Blumstengel}
\affiliation{Institut f\"ur Physik,
Humboldt--Universit\"at zu Berlin,
Newtonstra\ss{}e 15,
D-12489 Berlin, F. R. Germany}

\author{Fritz Henneberger}
\affiliation{Institut f\"ur Physik,
Humboldt--Universit\"at zu Berlin,
Newtonstra\ss{}e 15,
D-12489 Berlin, F. R. Germany}

\author{Volkhard May}
\affiliation{Institut f\"ur Physik,
Humboldt--Universit\"at zu Berlin,
Newtonstra\ss{}e 15,
D-12489 Berlin, F. R. Germany}
%

%
%
%


\begin{abstract}
We propose a new approach for calculating the change of the absorption spectrum of a molecule when moved from the gas phase to a crystalline morphology. 
The so-called gas-to-crystal shift $\Delta{\cal E}_m$ is mainly caused by dispersion effects and depends sensitively on the molecule's specific position in the nanoscopic 
 setting. 
Using an extended dipole approximation, we are able to divide $\Delta{\cal E}_m= -Q W_{m}$ in two factors where 
$Q$ depends only on the molecular species and accounts for all non-resonant electronic transitions contributing to the dispersion, while $W_m$ 
is a sum running over the position of all molecules expressing the site--dependence of the shift in a given molecular structure. The ability of our approach to predict absorption spectra
is demonstrated using the example of polycrystalline films of 3,4,9,10-perylene-tetracarboxylic-diimide (PTCDI).
\end{abstract}
\maketitle

Thin films of conjugated organic molecules are 
essential building blocks in organic and hybrid opto-electronics. These films are typically characterized by a 
nanocrystalline structure. 
Depending on the application, planar or bulk heterojunctions are realized either of different organic molecules or by combination with inorganic semiconductors or metal oxides such as ZnO and TiO$_2$ \cite{kamat07,umeyama09,Pneumans2003}.
The energetic alignment of the ground and excited states of the different materials governs the efficiency of charge separation and the open--circuit voltage in photovoltaic cells and, similarly, the injection of charge carriers and energy transfer dynamics in light-emitting devices \cite{Meyer2012,Koch2007,Kirchartz2009}. 
Hence,
it is essential to understand in detail how 
structural disorder and site-dependent dispersion affect the energetic properties of organic thin films.

Due to the large number of molecules involved, a full quantum mechanical treatment of the polycrystalline thin film is far out of reach for modern computational
electronic structure methods. 
Hence, many techniques used in quantum chemistry follow a different approach, in which only a single molecule is treated quantum mechanically and coupled to the environment,
which itself is treated classically (for a recent overview see \cite{Mennucci2013}). Consequently, these techniques
result in molecular wave functions and energies that 
parametrically
depend on the environmental coordinates,
which may again become
computationally expensive. 

Here, we will use a different approach, in which only the excited states of the isolated molecule are calculated. In a second step, these states are then coupled to the environment via coupling matrix elements calculated from a simplified Hamiltonian that takes into account the entire environment on a nano-scale. This methodology will be used in the following to calculate the shift of the single-molecule excitation energies due to the coupling to the specific inhomogeneous environment encountered
in nano--structured polycrystalline molecular films. We note, that it was shown in \cite{Gisslen2009} that charge transfer states coupling to Frenkel exciton states
changes the absorption curve only slightly for PTCDA crystals. Such a small effect was reproduced in our own calculations using the parameters given in \cite{Gisslen2009}.
For the sake of clarity we will neglect charge transfer states in this work.

Generally speaking,
when a molecule $m$ is moved from the gas phase into a crystalline morphology, its optical transition energies undergo the so--called gas--to--crystal shift.
This shift is due to (1.)~Coulomb coupling of static charge distributions, (2.)~inductive polarization shifts, (3.)~excitonic shifts, and (4.)~dispersion shifts, caused by the interaction
between higher excited transition densities of the molecule $m$ and surrounding molecules.
The effect of electrostatic couplings to the energy shifts was calculated as interaction of partial charge distributions (see e.g.~\cite{Madjet2006, Megow2014-2}, calculation details in \cite{remark2})
and is included in the diagonal elements ${\cal H}_{m m}$
of the standard Frenkel exciton Hamiltonian. Since the respective energy shift due to electrostatic Coupling (1.) was found to be small ($<$10 meV), we neglect the contribution of inductive polarization shifts (2.), which is
supposed to be in the same order of magnitude as (1.). Excitonic couplings (3.) are represented in the non--diagonal elements ${\cal H}_{m n}$ of the Frenkel exciton Hamiltonian.
The dispersive shift (4), which is very large for the most systems (up to several 100 meV), is often assumed to be constant for all molecules of a system. In this work we will derive a site--dependent dispersive
correction ${\Delta \cal E}_m$ to the diagonal Hamiltonian elements ${\cal H}_{m m}$.

\begin{figure}[t]
\centering
\includegraphics[width=0.4\textwidth]{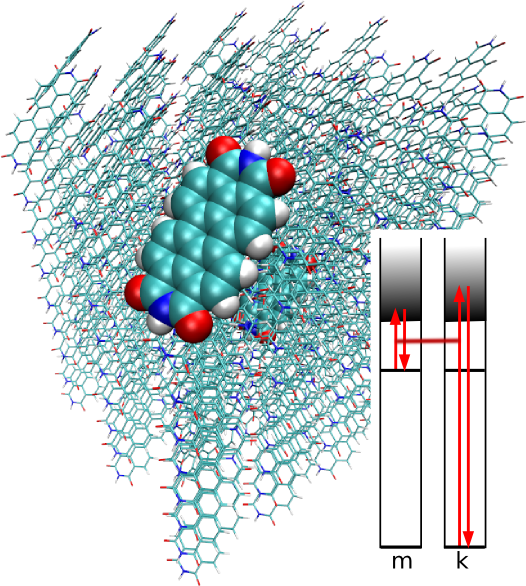}
\caption{PTCDI crystallite (cube of about 4.5 nm edge length with structure obtained 
from a 1 ns MD simulation). Highlighted are one PTCDI molecule at the crystallite surface
and one in the interior
(both molecules are shown via  space--filling models).
Lower right: scheme of coupled virtual transitions giving rise to the dispersive
shift $\Delta E_{m e}(k)$, Eq. \er{energy-shift}. Shown is the
transition of molecule $m$ into a higher level $f$ and the transition of the single
molecule $k$ of the surrounding from its ground--state to the higher level $f'$. 
All figures were produced with VMD \cite{VMD}.
}
\label{Fig1}
\end{figure}

\begin{figure}[t]
\centering
\includegraphics[width=0.48\textwidth]{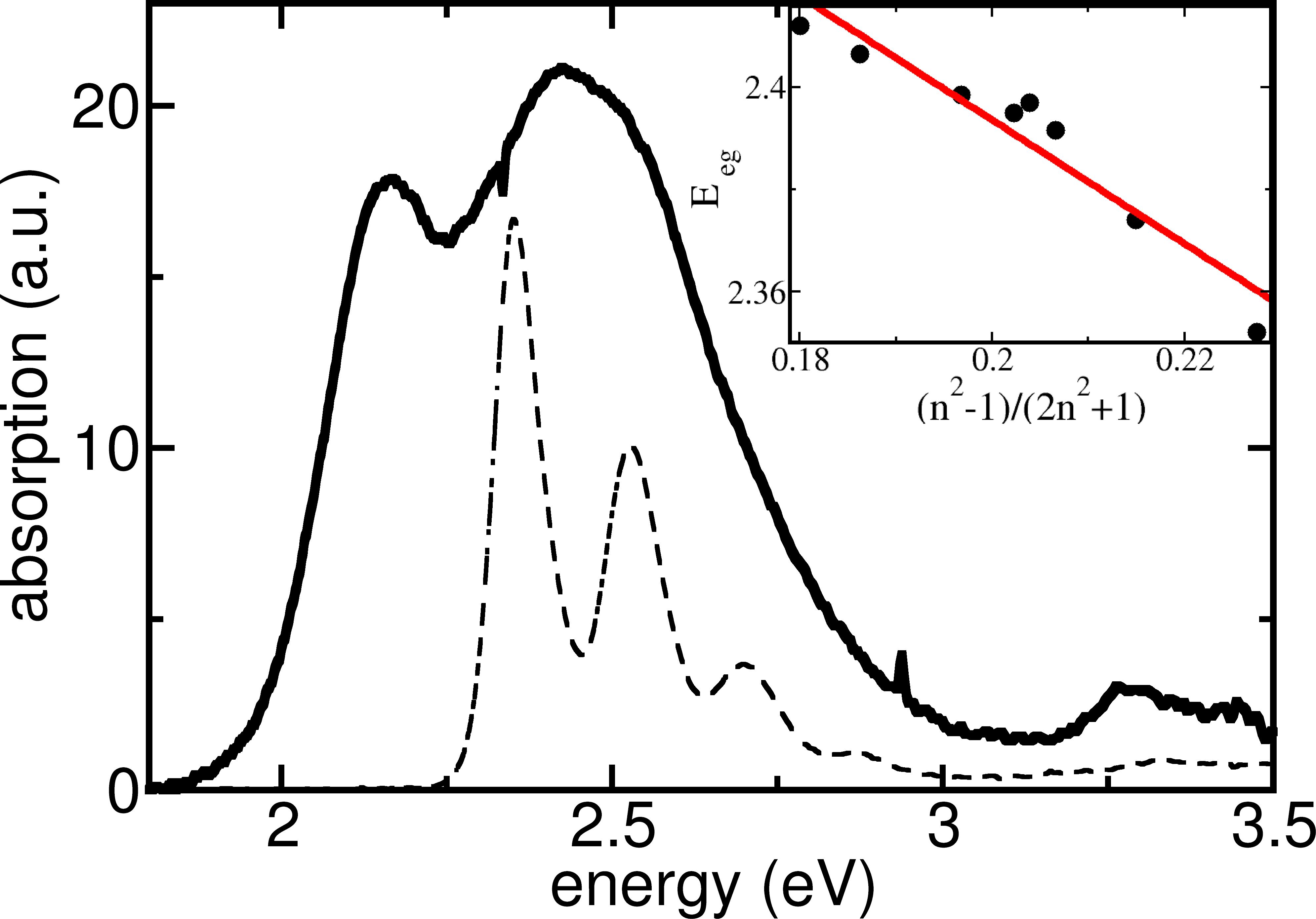}
\caption{Room ­temperature absorption line­shape of different PTCDI systems.
Solid black line: PTCDI film vacuum--deposited on Al$_2$O$_2$. The nominal thickness is 12 nm. Dashed black line: PTCDI monomer in toluene solution.
The inlay shows the $S_{0,\nu=0}\rightarrow S_{1,\nu=0}$ transition energy for the PTCDI derivate N,N'-Bis(1-hexylheptyl)-perylene-3,4:9,10-bis-(dicarboximide) in 
non--polar solvents with different refractive indices $n$: 
pentane ($n=1.36$), hexane ($n=1.38$), nonane ($n=1.41$), dodecane ($n=1.42$),
cyclohexane ($n=1.43$), hexadecane ($n=1.43$), tetrachlormethane ($n=1.46$), 
and benzene ($n=1.50$); red line: linear fit.
\\
}
\label{Fig2}
\end{figure}

While in a bulk crystal all molecules feel the same infinitely
large environment, the dispersion becomes site-dependent in a nanoscopic morphology.
In a finite crystallite, for example,
the gas--to--crystal shift will strongly depend on whether the molecule is situated close to the surface or deep within the crystallite (see Fig. \ref{Fig1}).
In this letter, we introduce an efficient computation scheme of the site--dependent level shift and demonstrate its capability for providing a consistent understanding of the optical absorption
spectra of thin nano--crystalline films of 3,4,9,10-perylene-tetracarboxylic-diimide (PTCDI, cf.~Fig.~\ref{Fig1}).
As a consequence of disorder and site-dependent dispersion, the thin film spectrum is markedly different from that of PTCDI in solution (cf.~Fig.~\ref{Fig2}). 
In particular, it exposes a distinct double structure that cannot be explained by features of the isolated molecule.
According to X-ray diffraction measurements, the film morphology is dominated by crystallites
with linear extension smaller than 15 nm and a random in-plane orientation. Hence, crystalline domains may consist of some hundreds of PTCDI molecules.

%
%

We start the description of the system by defining the electronic states $\varphi_{m a}$ and related energies $E_{m a}$ of an isolated molecule at site $m$ and in state $a$.
In the most naive picture, singly excited states now follow from the standard Frenkel--exciton Hamiltonian
$H_{\rm exc} = \sum_{m, n} {\cal H}_{m n} \ket{\phi_m}\bra{\phi_n}$, where the ground state energy $E_0 = \sum_m E_{m g} = 0$ is shifted to the origin of the energy scale.  
Here, we have defined the molecular product states $\phi_m$ with molecule $m$ in the first excited state $\varphi_{m e}$ and all other molecules in the
ground state $\varphi_{n g}$. The overall ground state $\phi_0$ is given by the product of all molecular ground-states $\varphi_{n g}$. The diagonal and off--diagonal part of the Hamiltonian matrix ${\cal H}_{m n}$ represent the excitation energies 
${\cal E}_m = E_{m e} - E_{m g}$ and the resonant excitonic coupling terms 
${\cal J}_{m n}$, respectively. A proper rescaling of all elements of
${\cal H}_{m n}$ 
may account for the dispersion effects, but a more microscopic picture is desirable. 



In the following, we will directly compute the site--dependent shifts $\Delta {\cal E}_m = \Delta E_{m e} - \Delta E_{m g}$ of the excitation energy ${\cal E}_m$. 
Here, the individual level shifts $\Delta {E}_{m a}$ with $a=g,e$ arise from the non-resonant Coulomb coupling of molecule $m$ with all its surrounding molecules. They can be written as $\Delta E_{m a} = \sum_k \Delta E_{m a}(k)$, where the 
contribution due to the coupling to molecule $k$ is  given in second order perturbation theory by 
\cite{Salam2010, Renger2008}
\bea
\label{energy-shift}
\Delta E_{m a}(k) = - \sum_{f, f'} 
\frac{|J_{m k}(a g, f' f)|^2}{E_{m f a} + E_{k f' g}} \; .
\eea
%
The expression includes the transition energies 
$E_{m f a} = E_{m f} - E_{m a}$ and $E_{k f' g} = E_{k f'} - E_{k g}$, 
$f$ and $f'$ count all higher excited energy levels
($f, f' > e$), and $J_{m k}(a g, f' f)$ is the Coulomb coupling between the electronic
transition density $\rho^{(m)}_{f a}({\bf r})$ of molecule $m$ and the electronic
transition density $\rho^{(k)}_{f' g}({\bf r'})$ of molecule $k$ (cf.~Fig.~\ref{Fig1}) \cite{Madjet2006}. Eq. \er{energy-shift}
illustrates that the site--dependence of the dispersion originates from the dependence of the $J_{m k}(a g, f' f)$ on the position of
molecule $m$ relative to molecule $k$. Since the energy difference in the denominator is always more than one order of magnitude smaller than the coupling of transition densities, Eq.~\ref{energy-shift}
represents a good approximation for the energy shift $\Delta E_{m a}(k)$.

The rigorous evaluation of Eq. \er{energy-shift} for a large molecule like PTCDI in a complex environment like 
a nano--crystalline film according to Eq. \er{energy-shift} is a formidable task. 
Instead, we utilize 
a simplified treatment based on the well-established extended dipole approximation for computation of the $J_{m k}(a g, f' f)$. This allows us to derive a formula 
for the level shifts that contains
only quantities which are, at least in principle, accessible by experiment.  
In the extended dipole approximation, the actual transition densities are replaced by transition dipoles represented by a single negative and a single
positive charge $\mp q$ placed 
at a fixed distance that resembles the spatial extension of the single-molecule excitation.
As a consequence,
the resulting distance dependence of the $J_{m k}(a g, f' f)$ is more realistic as in the point dipole approximation.

%
%

%
\begin{figure}
\centering
\includegraphics[width=0.25\textwidth]{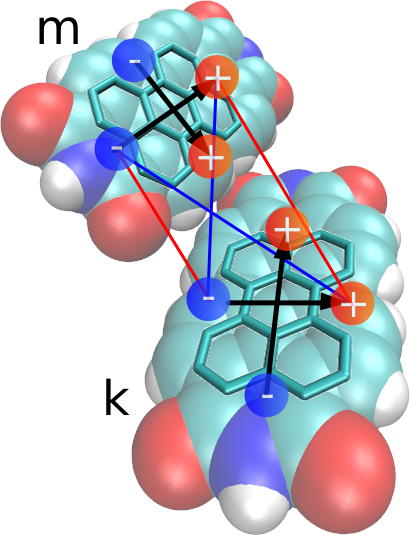}
\caption{Space--filling model of two PTCDI molecules. The carbon bondings
establishing the $\pi$--electron system are highlighted. Also shown are the positive and
negative partial charges
of an extended dipole model for higher electronic transitions of type $\xi$.
It results a position and orientation dependent excitonic coupling between
molecule $m$ (upper left) and molecule $k$ (lower right). The shown interaction
constitutes $V(m  \|, k \bot)$.
}
\label{Fig3}
\end{figure}
%
The PTCDI excited states can be classified into states with transition
dipoles parallel to the long molecular axis ($\xi(f) = \|$) and perpendicular to it ($\xi(f) = \bot$)
\cite{remark2}. 
The basic idea of our simplified treatment 
is to place the charges at the
same distance for all transitions $f$ with polarisation ($\xi(f) = \|$), and likewise for the charges corresponding to the transitions of type $\xi(f) = \bot$. In both cases, the $q(f)$ are positioned at the 
boundary
of the $\pi$-electron system (cf. Fig. \ref{Fig3}). 
As a consequence, the charge $q(f)$ depends
on the excited state $f$ and 
can be derived from
the ratio of the respective transition dipole moment and the intercharge distance. 

We verified this approximation by performing extensive electronic structure computations for a large number of excited states by comparing the Coulomb coupling matrix elements $J_{m k}$ obtained by the extended
dipole approximation with the exact value based on atomic-centered partial transition charges  \cite{remark2}. 

The mean error is below 5 \%
for all excited states that have been considered. 
The extended dipole approximation allows us to 
factorize the interaction matrix element
$J_{m k}(a g, f' f) = q(f) q(f') V(m \xi(f), k \xi'(f'))$ into two contributions, 
the effective charges $q(f)$, $q(f')$ and the distance dependence of the interaction of extended dipoles $V(m \xi(f), k \xi'(f'))$, i.~e.~the interaction
of a pair of unit charges of opposite sign at molecule
$m$ and $k$. The $V(m \xi(f), k \xi'(f'))$ still depend on the transitions $f$, $f'$ via $\xi(f)$, $\xi'(f')$.

If the different states $f$, $f'$ lie sufficiently close in energy,
we may average over the two different transition dipole directions by introducing
$V_{mk}=\sum_{\xi, \xi'} V(m\xi, k\xi')/4$. Our electronic structure calculations of the higher excited states support this approximation.
As a consequence,
the $V_{mk}$ become independent on the states $f$, $f'$ and hence $J_{mk}(ag,f'f)\approx q(f)q(f')V_{mk}$. 
Finally, the site--dependent transition energy shift of molecule $m$ can be written as 
\bea
\label{single-shift}
\Delta  {\cal E}_m  = -  Q W_m,
\eea
with 

\bea
&&
Q = \sum_{f, f'} q^2(f) q^2(f') 
\Big( \frac{1}{E_{m f e} + E_{k f' g}}
\nl
- \frac{1}{E_{m f g} + E_{k f' g}} \Big) \; 
\eea
and $W_m = \sum_k V_{m k}^2$.



While $W_m$ can be interpreted as a geometry factor determined by the structure of the nano--crystalline film, 
the $Q$--factor is specific to the molecule under consideration and accounts for transition energies and strengths of all its excited states $f$. 
As the $Q$-factor can not directly be calculated,
it can
be obtained from an independent absorption measurement of an amorphous film of randomly orientated molecules.
The obtained value of $\Delta {\cal E}_m$ is identified with
$Q \left\langle W_m\right\rangle$. The thus obtained $Q$--factor can then be used to calculate site-dependent energy shifts for any 
given nano-structured environment.

In the following, we apply this methodology to explain the optical absorption spectra of nano--crystalline PTCDI films (cf.~Fig.~\ref{Fig2}). 
It is important to note that both, the resonant excitonic coupling (which causes a mixing of states) as well as the non--resonant interactions 
(which are responsible for the level shifts) contribute to the observed changes in the spectrum when going from the isolated molecule to the polycrystalline thin film.
Before simulating PTCDI absorption spectra, we provide an experimental estimate of the expected magnitude of the level shift. To this end,
we measured absorption spectra placing the molecules in various non--polar environments (solvents). Since PTCDI is not well soluble, we perform the 
experiment with the related molecule N,N'-Bis(1-hexylheptyl)-perylene-3,4:9,10-bis-(dicarboximide) whose UV/VIS spectrum perfectly coincides with that
of PTCDI as the alkyl substituents do not affect the $\pi$-electron system. The shift of the transition energy $\Delta {\cal E}_m$ with respect to the
gas phase excitation energy $E_{\rm gp}$ is related to the solvent's refractive index $n$ at optical wavelengths via  
$\Delta {\cal E}_m = - {\cal F} f(n)$ with  $f(n) = (n^2 - 1)/(2n^2 + 1)$ \cite{Bayliss1950, Renger2008}. The expression originates from the solvation
energy of a point dipole in a spherical cavity surrounded by a homogeneous continuum \cite{Bayliss1950}. We will refer to this relation
as the continuum solvent approximation (CSA). A fit of the measured $S_{0,\nu=0}\rightarrow S_{1,\nu=0}$ transition energy as a function of the solvent's
refractive index (cf. inset of Fig.~\ref{Fig2}) yields ${\cal F} = 1.21$ eV and $E_{\rm gp} = 2.64$ eV. For a thin polycrystalline 
PTCDI film, a refractive index $n_{\rm PTCDI} = 1.96$ is reported \cite{El-Nahhas2012, remark5}. Application of CSA predicts a rather substantial
level shift of $\Delta {\cal E}_m = - 400$ meV with respect to $E_{\rm gp}$ 
which is in the range of the observed spectral changes (cf.~Fig.~\ref{Fig2}). 

 
In order to calculate the $W_m$ and the excitonic coupling matrix elements ${\cal J}_{m n}$, a precise knowledge of the morphology (size distribution
and orientation of grains) and arrangement of PTCDI molecules in nano--crystalline domains is required. However, the experimental determination of the exact 
structure of the present PTCDI film is beyond the scope of this Letter.  Instead,  we performed extensive MD simulations of a set of 216 PTCDI molecules forming cubic
crystallites of maximal edge length of 4.5 nm in order to obtain possible PTCDI crystallite configurations \cite{remark3}. 

For a quantitative derivation of absorption spectra, three more factors have to be accounted for, that is, the computation of ${\cal J}_{m n}$,
the screening of ${\cal J}_{m n}$ and the vibronic coupling. 
Here, the ${\cal J}_{m n}$ were caluclated using the molecular coordinates and transition partial charges \cite{Madjet2006}.
Screening can be treated with the so--called
{\it Poisson--transition--charges--from--electrostatic--potential}--method 
\cite{Adolphs2006} or with methods based on the {\it polarizable continuum
model} and beyond \cite{Curutchet2011}. Recently, 
some of us 
demonstrated that a $1/\epsilon$--screening approach, in which ${\cal J}_{m n}$ is replaced by  ${\cal J}_{m n}/\epsilon$, is sufficient as long as only ensemble averages are calculated \cite{Megow2014-2}. Hence, the same approach was used here, yielding maximal values of ${\cal J}_{m n}/\epsilon \approx 20$ meV for the MD derived geometry. 


%
\begin{figure}
\centering
\includegraphics[width=0.4\textwidth]{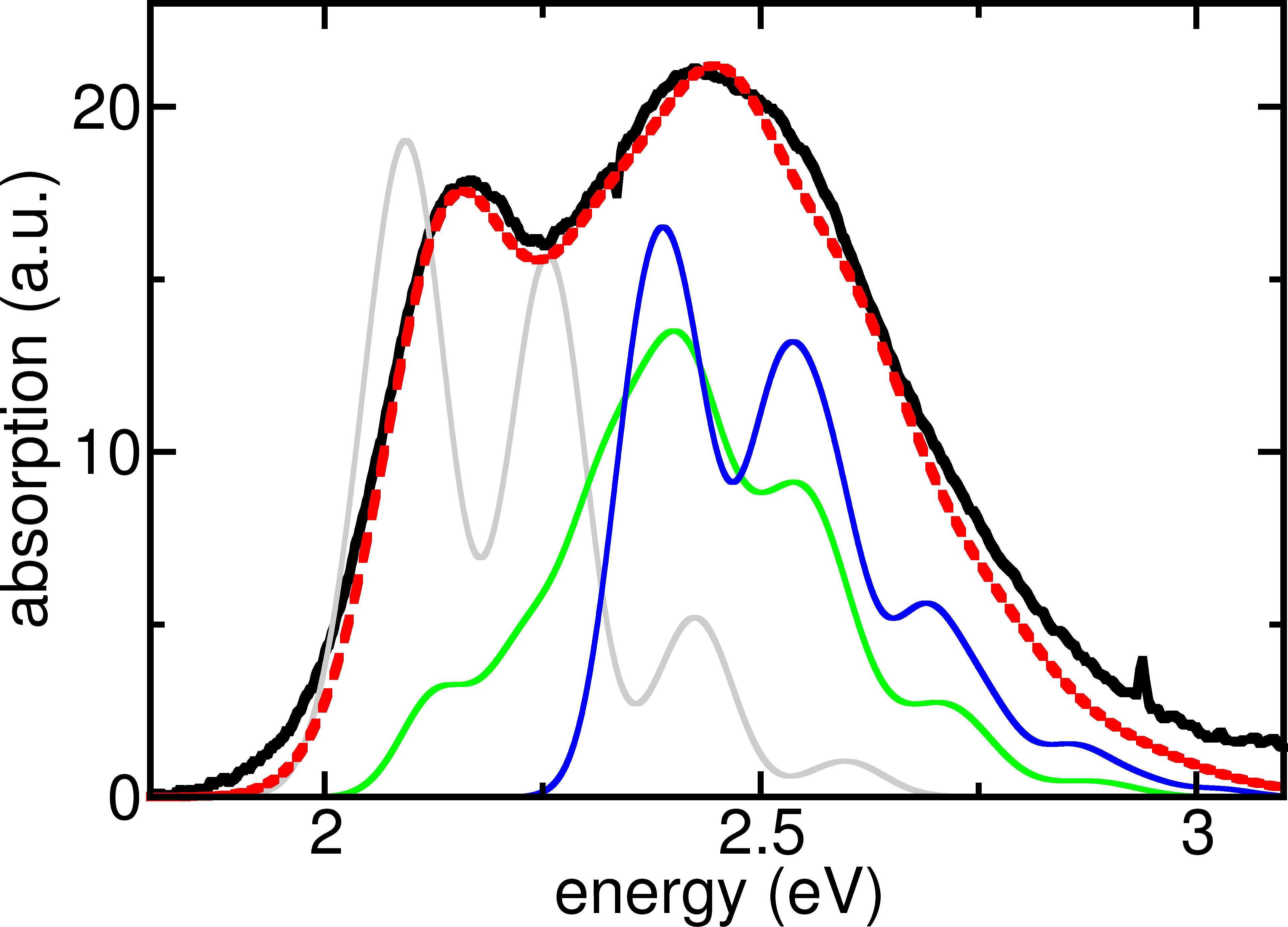}
\caption{Computed absorption line­shapes of different types
of PTCDI nano­crystalline films in comparison to the experimental absorption spectrum (black curve). Grey curve: large cubic crystallites with 4.5 nm edge length;
green curve: medium-size crystallites (48 molecules); blue curve: very small crystallites (10 molecules);  dashed red curve: 
theoretical absorption spectrum calculated from a mixture of very small and large crystallites (see text for details).
 }
\label{Fig4}
\end{figure}
%

To account for vibronic coupling, only a single vibrational progression per molecule (Huang-Rhys-factor: 0.62) is taken into account. This has been found sufficient for 
modelling the PTCDI--monomer spectrum \cite{Scholz2006}. The single particle approximation \cite{Spano2002} is used to derive an electron--vibrational Hamilton 
matrix where a single vibronically (labelled by $\nu$) excited molecule couples to surrounding molecules in the vibrational ground state.
Exciton states are constructed as
$\Phi_\alpha = \sum_m c_\alpha (m, \nu) \chi_{m e \nu} \phi_m$, where $\chi_{m e \nu}$ is the total
vibrational wave function with vibrational excitations in the excited state of molecule
$m$ and the $c_\alpha (m, \nu)$
denote the respective expansion coefficients.  The absorption line--shape follows as the 
sum over all exciton levels $\alpha$ with transition energies $E_\alpha$ weighted by the respective oscillator strengths $|\sum_{m, \nu} c_\alpha (m, \nu) d_m |^2$. 
Finally, Gaussian broadening of typically 0.06 eV 
(full width at half maximum)  
is introduced for all transitions. The value is deduced from the experimental absorption line width of the monomer in solution.

As noted above, resonant excitonic coupling is rather weak in PTCDI nano--crystallites. Therefore, site-dependent 
level shifts must be mainly responsible for the observed spectral changes (cf. Fig. \ref{Fig2}). The thin film spectrum
is composed of a narrow, red-shifted absorption peak and a broad band in the spectral range of the solution spectrum. 
This line--shape suggests that the spectrum is a superposition of spectra of a densely packed crystalline phase and a more 
loosely packed rather disordered phase. Therefore, we consider two distinct types of nano--scaled crystallites: First, we take the 
periodical PTCDI cube of 4.5 nm edge length derived from MD simulations and evaluate $W_m$ for all molecules. To obtain the 
absorption spectrum, the Hamiltonian for the molecules contained in one cube is diagonalised. As the exciton spectrum does not 
further change when considering larger systems, the obtained spectrum is representative for films composed of crystallites of 
this size and larger. A more loosely packed disordered phase is modelled by cutting very small blocks consisting of 10 molecules
out of the 4.5 nm cube. The distance in $x$, $y$ and $z$-direction between van--der--Waals--shells of neighbored molecules that are positioned in different blocks is set to 2.5 \AA{ }
reducing the packing density of films
composed of these very small crystallites by 30 \%. Of course this kind of model system does not include isotropic orientations of the crystallites, which are likely present in the experiment.
But, it includes larger distances between different crystallites due to the smaller packing density, which strongly affect the energy shift for molecules that are positioned at the border of a crystallite.
Absorption spectra are calculated
setting $Q = 4.26$ eV \AA$^2$. This value follows from our assumption concerning the size--distribution of crystallites in the film which is based on
the spectral position and amplitude of the experimental absorption features of
the thin film with respect to $E_{\rm gp}$ (cf.~Fig.~\ref{Fig4}). 



The shape of the absorption spectra calculated for both the densely packed crystalline film and the
loosely packed disordered film resembles that of PTCDI in solution. As anticipated above, excitonic coupling does indeed not 
noticeably alter the shape of the spectrum. Furthermore, in the film composed of solely large crystallites, most molecules are 
located in the interior of a crystallite and, thus, experience a similar level shift. The same holds for the film of the very small 
crystallites where all molecules are close to a surface with an accordingly smaller level shift. The possible modification of the 
absorption spectrum by the site dependence of the level shift becomes more apparent in the intermediate case of a film composed of 
medium size nano--crystallites containing 48 molecules where both molecules in the interior and on the surface noticeable contribute 
to the spectrum (cf. Fig. \ref{Fig4}). Here, the mean distance between crystallites is set to 5 \AA{ } to maintain the packing density of
the loosely packed film. Finally, to simulate the experimental absorption spectrum, the average is taken over a mixture of very
small and large nano-crystallites. In order to take into account the larger disorder in the very small crystallites, the
inhomogeneous width is increased to 0.20 eV.
As a result, agreement with the experimental spectrum is obtained when assuming that the PTCDI film consists to 51 \% of the large and to 49 \% of the very small crystallites (cf. Fig. \ref{Fig4}).  
It has to be noted that an agreement of experimental and theoretical spectrum can only be obtained, if the two respective films inhibit energy shifts with a difference of about 0.3 eV. Noting that the $Q$ value will always 
be limited to
$1.8 \text{ eV \AA$^2$} < Q < 5.9$ eV \AA$^2$ (results for completely loosely packed and completely dense film), an energy shift of 0.3 eV can only be achieved using a combination of large and very small crystallites.
A change of the $Q$-parameters within the given limits shifts the whole spectrum, while its effect on the energy shift difference is comparable small. This illustrates that the given formalism allows only a small amount
of fitting.
%

%
In summary, we introduced a new approach for calculating the optical absorption spectra of organic polycrystalline thin films. In particular,
a novel relation for molecular excitation energy shifts due to dispersion effects of the environment has been derived. The obtained formulas allow for the
determination of site--dependent level shifts of a molecule in a given nanoscale environment. 
On that base, we were able to explain a distinct double structure of the S$_0\rightarrow$S$_1$
transition in the absorption spectrum of PDCTI films by the coexistence of two
molecular phases. Large crystals where most molecules are located in the interior
give rise to the component on the low-energy side. The second feature is originating
from a phase of loosely-packed small aggregates resulting in a weaker gas-to-crystal
shift as most molecules reside at or close to the surface. While a fully quantitative
analysis of the film morphology is beyond the scope of this work, our findings
emphasize that explicite account of the local energy structure is mandatory for
understanding the optical properties of molecules in nano-scaled solid-state
systems.

%
{\it Acknowledgments:}
%
Financial support by the {\it Deutsche Forschungsgemeinschaft} through Sfb 951
and through project ME 4215/2-1 (JM) and by the Austrian Science Fund (FWF): P24774-N27 (TR) is gratefully acknowledged. We are indebted to Frank M\"uh (JKU Linz) and Stefan Kowarik (HU Berlin) for
fruitful discussions.

%

\begin{references}
%
%


  %
  \bibitem{kamat07} Kamat,~P.~V.
  \textit{J.~Chem.~Phys.~C}  \textbf{2007,} \textsl{111,} 2834-2860. 
%
 
  \bibitem{umeyama09}Imahori,~H.;\ \ Umeyama,~T.
  \textit{J.~Chem.~Phys.~C}  \textbf{2009,} \textsl{113,} 9029-9039. 
  
  \bibitem{Pneumans2003} Peumans,~P;\ \ Uchida,~S;\ \ Forrest,~S.~R. \textit{Nature} \textbf{2003,} \textsl{425,} 158.
  
  \bibitem{Meyer2012} Meyer,~J.;\ \ Hamwi,~S.;\ \ Kr\"{o}ger,~M.;\ \ Kowalsky,~W.;\ \ Riedl,~T.;\ \
Kahn,~A. \textit{Adv.~Mater.}~\textbf{2012,} \textsl{24,} 5408.

  \bibitem{Koch2007} Koch,~N. \textit{ChemPhysChem} \textbf{2007,} \textsl{8,} 1438.

  \bibitem{Kirchartz2009}Kirchartz,~T.;\ \ Taretto,~K.;\ \ Rau,~U. \textit{J.~Phys.~Chem.~C} \textbf{2009,} \textsl{113}, 17958.

\bibitem{Mennucci2013} Mennucci,~B \textit{Phys.~Chem.~Chem.~Phys.}~{\bf 2013,} \textsl{15}, 6583.
  
%


\bibitem{Gisslen2009}
Gissl\'en,~L.;\ \ Scholz,~R.
\textit{Phys. Rev. B}~ \textbf{2009,}, \textsl{80}, 115309.

\bibitem{Madjet2006}
Madjet,~M.~E.-A.;\ \ Abdurahman,~A.;\ \ Renger,~T. \textit{J.~Phys.~Chem.~B}
  \textbf{2006,} \textsl{110,} 17268-17281. 
  
 \bibitem{Megow2014-2}
Megow,~J.;\ \ Renger,~T.;\ \ May,~V. \textit{ChemPhysChem} \textbf{2014,} \textsl{15}. %

\bibitem{remark2} The mentioned classification of excited state is the
outcome of considerable amount of electronic structure computations on a single PTCDI
molecule. DFT and TDDFT calculations were performed in utilizing  the
Gaussian 09 (Rev. A) program \cite{Frisch2009}, the B3LYP functional
\cite{SDCF94} and a 6-311G(d,p) basis-set.
The optimized ground-state geometry of PTCDI was verified using frequency
analysis. Transition partial charges were calculated using TDDFT followed by a
natural transition orbital (NTO) analysis \cite{Ma03} .
  
  \bibitem{VMD}
Humphrey,~W.;\ \ Dalke,~A.;\ \ Schulten,~K.
{VMD} - {V}isual {M}olecular {D}ynamics.
J.~Mol.~Graph.~ \textbf{1996,} \textsl{14}, 33-38. 
  

\bibitem{Salam2010}
\textit{Molecular quantum electrodynamics :
long-range intermolecular interactions}, Hoboken, N.J. :
Wiley \textbf{2010,} Formerly CIP.


\bibitem{Renger2008}
Renger,~T.;\ \ Grundk{\"{o}}tter,~B.;\ \ Madjet,~M.~E.-A.;\ \ M{\"{u}}h,~F.
\textit{Proc.~Natl.~Acad.~Sci.}~\textbf{2008,} \textsl{105,} 13235-13240. 








   
\bibitem{Bayliss1950}
Bayliss,~N.~S \textit{J.~Chem.~Phys.}~ \textbf{1950,} \textsl{18,} 292-296. 




\bibitem{El-Nahhas2012}
El-Nahhas,~M.~M.;\ \ Abdel-Khalek,~H.;\ \ Salem,~E. \textit{Adv.~Condens.~Matter Phys.}~ \textbf{2012,} \textsl{2012}.

\bibitem{remark5} 
We assume that the crystallite distribution in \cite{El-Nahhas2012} for which the refractive index $n_{\rm PTCDI}=1.96$ was 
deduced is similar to our film since the two absorption spectra are nearly similar.


\bibitem{remark3}MD simulations  have been carried in
using NAMD program package \cite{Phillips2005}, the AMBER force field
\cite{Case2004} and the GAFF parameter sets \cite{Wang2004-2}.
A set of 216 PTCDI molecules was considered forming a cubic crystallite of
4.5 nm edge
length and with
two molecules in the unit cell \cite{Klebe1989}. Moreover, periodic boundary
conditions have been established and
the electrostatic interactions were computed using the particle mesh Ewald
method
\cite{Darden1993}.
The temperature was increased to 300 K, and 1 ns simulations were carried
out. Since
no spectral changes could be observed after 0.5 ns of MD simulation,
the obtained MD data was considered to be sufficient.

\bibitem{Adolphs2006}
Adolphs,~J.;\ \ Renger,~T. \textit{Biophys.~J.}~ \textbf{2006,} \textsl{91,} 2778-2797. 



 \bibitem{Curutchet2011}
Curutchet,~C.;\ \ Kongsted,~J.;\ \ Mu{\~{n}}oz-Losa,~A.;\ \ Hossein-Nejad,~H.;\ \ Scholes,~G.~D.;\ \ Mennucci,~B. \textit{J.~Am.~Chem.~Soc.}~ \textbf{2011,} \textsl{133,} 3078-3084. 
  


\bibitem{Scholz2006}
Scholz,~R.;\ \ Schreiber,~M \textit{Chem.~Phys.}~ \textbf{2006,} \textsl{325,} 9. 


\bibitem{Spano2002}
Spano,~F.~C.\textit{J.~Chem.~Phys.}~ \textbf{2002,} \textsl{116}. %



\bibitem{Frisch2009}
Frisch,~M.~J.  {\it et al.},
Gaussian~09 {R}evision {A}.1 \textbf{2009,}
Gaussian Inc. Wallingford CT 2009.

\bibitem{SDCF94}
 Stephens,~P.~J.;\ \ Devlin,~F.~J.;\ \ Chabalowski,~C.~F.;\ \ Frisch,~M.~J.
\textit{J.~Phys.~Chem.}~ \textbf{1994,} \textsl{98,} 11623. 

\bibitem{Ma03}
Martin,~R.~L.
\textit{J.~Chem.~Phys.}~ \textbf{2003,} \textsl{118}, 4775. 


\bibitem{Phillips2005}
Phillips,~J.~C.;\ \ Braun,~R.;\ \ Wang,~W.;\ \ Gumbart,~J.;\ \ Tajkhorshid,~E.;\ \
  Villa,~E.;\ \ Chipot,~C.;\ \ Skeel,~R.~D.;\ \ Kale,~L.;\ \
  Schulten,~K.
\textit{J.~Comput.~Chem.}~ \textbf{2005,} \textsl{26,} 1781-1802. 

\bibitem{Case2004}
Case,~D.~A. \textit{et al.},
Amber \textbf{,} \textsl{8} (2004),
University of California, San Francisco, CA.

\bibitem{Wang2004-2}
Wang,~J.;\ \ Wolf,~R.~M.;\ \ Caldwell,~J.~W.;\ \ Kollman,~P.~A.;\ \ Case,~D.~A.
\textit{J.~Comput.~Chem.}~ \textbf{2004,} \textsl{25,} 1157-1174. 


\bibitem{Klebe1989}
Klebe,~G.;\ \ Graser,~F.;\ \ H{\"{a}}dicke,~E.;\ \ Berndt,~J.
\textit{Acta Crystallog.~Sect.~B} \textbf{1989,} \textsl{45,} 69-77. 

\bibitem{Darden1993}
Darden,~T.;\ \ York,~D.;\ \ Pedersen,~L.
\textit{J.~Chem.~Phys.}~ \textbf{1993,} \textsl{98,} 10089-10092. 


%
\end{references}
%
%

%

%
%

%
\end{document}